\documentclass{iau}
\usepackage{graphicx}
\usepackage{bm}
\usepackage{wrapfig}

\title[Optical Design of Arago] 
{Preliminary design of the full-Stokes UV and visible spectropolarimeter for UVMag/Arago}

\author[M. Pertenais]   
{Martin Pertenais$^{1,2}$, Coralie Neiner$^1$, Laurent Par\`es$^{2,3}$, Pascal Petit$^{2,3}$, Frans Snik$^4$ \and Gerard van Harten$^4$}

\affiliation{$^1$LESIA, Observatoire de Paris, CNRS UMR 8109, UPMC, Universit\'e de Paris-Diderot, \\5 place Jules Janssen, 92100 Meudon, France \\ 
email: {martin.pertenais@irap.omp.eu} \\[\affilskip]
$^2$Universit\'e de Toulouse; UPS-OMP; IRAP Toulouse, France \\[\affilskip]
$^3$CNRS; IRAP ; 14 avenue Edouard Belin, 31400 Toulouse, France \\[\affilskip]
$^4$Leiden Observatory, Leiden University, P.O. Box 9513, 2300 RA Leiden, The Netherlands \\[\affilskip]
 } 

\pubyear{2015}
\volume{305}  
\jname{Polarimetry: From the Sun to Stars and Stellar Environments}
\editors{K.N. Nagendra, S. Bagnulo, \\ R. Centeno, \& M. Mart\'inez Gonz\'alez, eds.}
\begin{document}

\maketitle

\begin{abstract}
The UVMag consortium proposed the space mission project Arago to ESA at its M4 call. It is dedicated to the study of the dynamic 3D environment of stars and planets. This space mission will be equipped with a high-resolution spectropolarimeter working from 119 to 888 nm. A preliminary optical design of the whole instrument has been prepared and is presented here. 
The design consists of the telescope, the instrument itself, and the focusing optics. Considering not only the scientific requirements, but also the cost and size constraints to fit a M-size mission, the telescope has a 1.3 m diameter primary mirror and is a classical Cassegrain-type telescope that allows a polarization-free focus.  The polarimeter is placed at this Cassegrain focus. This is the key element of the mission and the most challenging to be designed. The main challenge lies in the huge spectral range offered by the instrument; the polarimeter has to deliver the full Stokes vector with a high precision from the FUV (119 nm) to the NIR (888 nm). The polarimeter module is then followed by a high-resolution echelle-spectrometer achieving a resolution of 35000 in the visible range and 25000 in the UV. The two channels are separated after the echelle grating, allowing a specific cross-dispersion and focusing optics for the UV and visible ranges. Considering the large field of view and the high numerical aperture, the focusing optic for both the UV and visible channels is a Three-Mirror-Anastigmat (TMA) telescope, in order to focus the various wavelengths and many orders onto the detectors.

\end{abstract}

\firstsection 

\section{Introduction}
\label{intro}
The Arago space mission is an ambitious project with a 1.3 m diameter telescope dedicated to spectropolarimetry in the UV and in the visible. 
 The measurement of spectra of all types of stars in the UV and visible domains provides important insights into their formation and evolution. The UV domain is very rich in atomic and molecular lines, it also contains the signatures of the stellar environment (e.g. of the chromosphere) and most of the flux of hot stars. At the same time, the visible spectrum allows us to gain information about the surface of the star itself. For the first time, 3D maps of stars and their environment will be reconstructed thanks to simultaneous UV and visible spectroscopy over a full stellar rotation period. The capabilities of extracting information on stellar magnetospheres, winds, disks, and magnetic fields will be multiplied tenfold by adding polarimetric power to the spectrograph. 
Arago will also be able to measure the changes in the stellar UV radiation and magnetic activity, allowing the study of the interaction between stars and their planets, in particular magnetospheric interactions and tides. The environmental conditions for the emergence of life on exoplanets will be characterized through these measurements thanks to Arago. 

Considering these different science cases, we are designing Arago with specific requirements on the performances of the instrument. The requirements and the detail of the instrument structure of Arago are presented in this article.

\section{Global Instrument and Specifications}
\label{spec}
\begin{figure}[h]
\begin{center}
\includegraphics[scale=0.33]{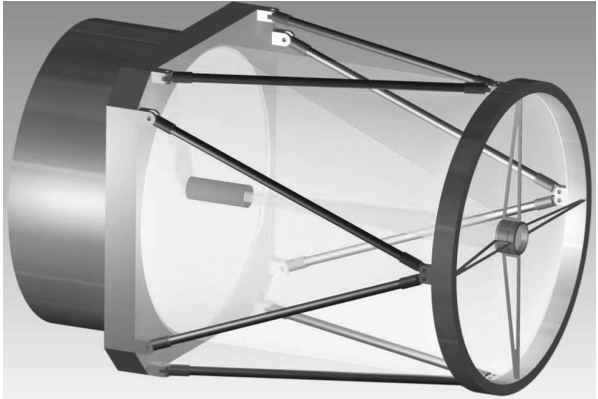}
\caption{\label{telescope-view} Global view of Arago's telescope and its instrument.
}
\end{center}
\end{figure}

The proposed mission includes a single instrument, placed at the Cassegrain focus of a 1.3 m telescope (see Figure~\ref{telescope-view}). To reach our scientific objectives, some specifications are defined for the instrument.

A wide spectral range is one of the key characteristics of the Arago mission. The requirement on this spectral range is [119:888] nm. The two extreme wavelengths were chosen to be able to observe the Ly$\alpha$ line in the FUV and important near-IR lines such as the triplet lines of calcium and the first Paschen hydrogen line.
The requirement of signal to noise ratio (SNR) expected is 100, calculated for a typical exposure time of 30 min for OBA stars of magnitude V=7 and 1h for FGK stars (V=7). This leads to a 1.3 m diameter telescope. The overall targets are, however, stars between V=3 and V=10. Using Least Squares Deconvolution (LSD) techniques (see \cite [Donati et al. 1997]{donati97}), the SNR ratio can be increased by a typical factor of 25.
We aim at high-resolution spectropolarimetry with Arago. The specification for spectral resolution is 35000 in the visible part of the spectrum and 25000 in the UV parts. These values were determined considering the spectral resolution needed by the various science cases. The spectral resolution is defined by the ratio of the wavelength over the resolved spectral element (over 2 pixels). For example, with Arago we will be able to resolve a spectral element of 0.015 nm at $\lambda$=525 nm or 0.005 nm at 125 nm.
Another very important concept for high-resolution spectropolarimeter is the accuracy of the radial velocity measurement. It is directly linked to the thermo-mechanical stability of the spectrometer, to the precision obtained in the positioning of spectral lines on the detectors and to the stability of these positions. Considering the absolute accuracy, the specification is V$_{\rm rad}$=1 km.s$^{-1}$. An even more important value is the relative accuracy of the radial velocity, during a measurement of a magnetic field. The specification is V$_{\rm rad}$=0.1 km.s$^{-1}$.
Finally, the last specifications for the instrument are related to the polarimetry. The science cases lead to a requirement in the polarization measurement error of less than 3\% and polarization sensitivity of 10$^{-3}$. Indeed, the polarization sensitivity will affect the capability to measure weak polarization information. On the other hand, accuracy of the polarization measurement is also an important issue. The challenge is to minimize every modification of the polarization state of the light by the optics on the optical path before the polarization analyzer. In particular, a close look has to be given to the polarization cross-talk effect between Q, U and V, and to the instrumental polarization induced by the instrument.
All these specifications are summarized in Table \ref{tab-spec} below.

\begin{table}[!htb]
  \begin{center}
  \caption{\label{tab-spec} Requirement details for the Arago instrument.}
 {\scriptsize
  \begin{tabular}{| l | c |}
\hline
{\bf Specification} & {\bf Minimum requirement} \\
 \hline
Spectral range & $ [119;320]$ nm + $[390;888]$ nm   \\ 
\hline
SNR & $100$  \\ 
\hline
Typical exposure time & 30 min (OBA stars, $V=7)$ \\
 & 1h (FGK stars, $V=7$)  \\ 
\hline
Target magnitude & $V=3-10$  \\ 
\hline
 UV Resolution & 25000  \\ 
\hline
Visible Resolution &35000  \\
 \hline
V$_{rad}$ accuracy & 1 km.s$^{-1}$ (absolute) \\
 & 0.1 km.s$^{-1}$ (relative) \\ 
\hline
Polarization accuracy & $<3\%$  \\ 
\hline
Polarization sensitivity & $10^{-3}$  \\ 
\hline
  \end{tabular}
  }
 \end{center}
\end{table}
\vspace{-8mm}
\section{Spectropolarimeter}
\label{spectropolarimeter}

	\subsection{Polarimeter}
Directly placed at the Cassegrain focus of the telescope (F/13 aperture) to avoid instrumental polarization (at first order), the polarimeter module is the most challenging component of the Arago payload. Indeed, the wide spectral range prevents us from using usual polarimeter designs as the ones from existing ground based spectropolarimeters, such as ESPaDOnS at CFHT and Narval at TBL at the Pic du Midi (see \cite[Auriere et al. 2003]{auriere03} and \cite[Donati et al. 2006]{donati06}).

The baseline chosen to develop a precise polarimeter on this wide spectrum is to use a temporal modulation of the polarization with optimal Stokes vector extraction efficiencies over the wavelength range, as proposed by \cite [del Toro Iniesta et al 2000]{deltoroiniesta00} and detailled by \cite [Tomczyk et al. 2010]{tomczyk10}.
To do so, we first define the total Mueller matrix M$_{tot}$ of the polarimeter that links the input and output Stokes vector S=(I Q U V)$^T$  by:  
\begin{equation}
S_{out}=M_{tot}.S_{in}
\end{equation}

\begin{figure}[h]
\begin{center}
\includegraphics[scale=0.25]{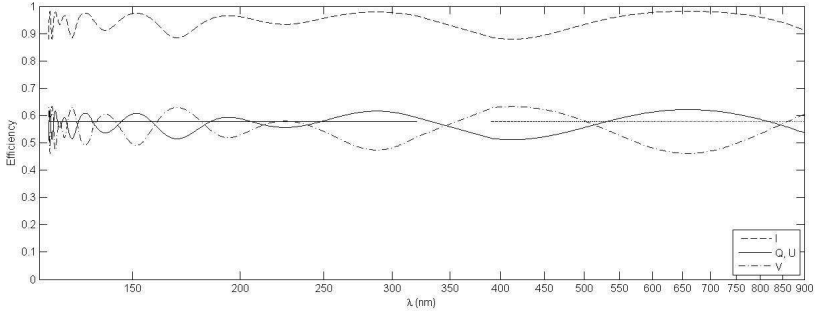}
\caption{\label{optimisation} The optimal extraction efficiencies for the Stokes parameters are close to the theoretical optimum : $57.7\%$ for Q, U and V, and $100\%$ for I.
}
\end{center}
\end{figure}

As the detector is only sensitive to the intensity, the only useful part is the first row of the Mueller matrix M$_{tot}$.The modulator rotates at 6 different angular positions, creating a temporal modulation of 6 different polarization states. We build the $6 \times 4$ modulation matrix O with the first rows of the total Mueller matrices M$_{tot}$ of the 6 states. Finally, a demodulation matrix D is defined, as detailed in \cite [Snik et al. 2012]{snik12} and its components are used to define extraction efficiencies of the Stokes parameters, $\epsilon_i$ with i=[1,2,3,4] for [I, Q, U, V]:
\begin{equation}
\epsilon_i={\left( 6 \cdot \sum_{j=1}^{6} D_{ij}^2 \right)}^{-1/2}
\end{equation}
The polarization modulator for Arago is built with 4 birefringent plates of Magnesium Fluoride (MgF$_2$), as it is the only transparent and birefringent material in the UV and visible domains. The goal is to optimize these plates (in thicknesses and relative fast-axis angles) along the spectral range, so that we will be able to extract the full-Stokes vector for every wavelength of interest. The two conditions on the efficiencies   and  tell implicitly that the optimal modulation efficiency is 100\% for Stokes I and $\frac{1}{\sqrt{3}}\approx57.7\%$  for Stokes Q, U and V. The 8 variables (4 thicknesses or retardance values and 4 fast axis orientations) are then numerically optimized over the whole spectral range to these ideal extraction efficiencies. This is shown in Figure~\ref{optimisation}.

\begin{figure}[h]
\begin{center}
\includegraphics[scale=0.13]{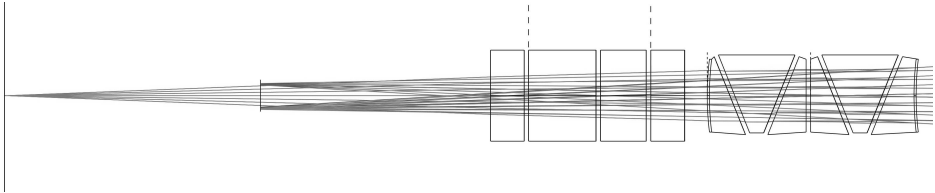}
\caption{\label{polarimetre} View of the preliminary design of the polarimeter. The stack of 4 MgF$_2$ retardance plates is followed by two 3-prims Wollaston beam-splitters. The various beams represent the two orthogonal polarization states.}
\end{center}
\end{figure}

To be able to measure the polarization modulation, the modulator has to be followed by a polarization analyzer, a so-called polarizer. A polarization beam-splitter is used to measure both orthogonal polarization states, to avoid losing flux and to eliminate every spurious effects in the measurement (e.g. imperfect flat-fielding, thermal drifts, or pointing jitter).
As the birefringence of MgF$_2$ is dramatically dropping in the FUV (see \cite [Pertenais et al. 2014]{pertenais14}), we need to use two 3-prims Wollaston beam-splitters. Otherwise, it would be impossible to separate the two orthogonal polarization states in the FUV. The preliminary design of the polarimeter is shown in Figure~\ref{polarimetre}.

The two 3-prisms Wollaston create a virtual slit (optically placed before the entrance of the modulator), which corresponds to the entrance slit of the echelle-spectrometer (see Section 3.2).
To compensate the axial chromatism created by the polarimeter and to maintain an output beam aperture of F/13, the entrance and exit faces of the polarization analyzer are spherically curved.

	\subsection{Spectrometer}
\begin{figure}[h]
\begin{center}
\includegraphics[scale=0.38]{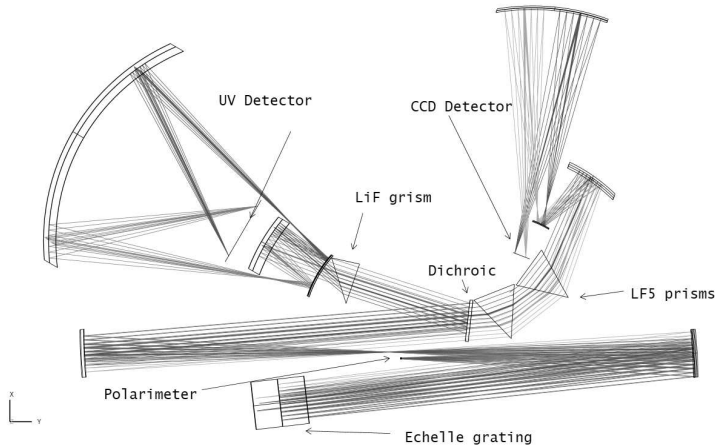}
\caption{\label{spectro} Preliminary optical design of the high-resolution echelle spectrometer. A single echelle grating is used to diffract all the wavelengths. A dichroic separates the UV from the visible part of the spectrum, before the cross-dispersers and focusing optics.}
\end{center}
\end{figure}

The second challenge in the development of the instrument for Arago is the creation of a single high-resolution spectrometer for this wide spectral range. The only viable solution is the use of an echelle-spectrometer. The preliminary design of this spectrometer is presented in Figure~\ref{spectro}.

As explained in the previous section, the entrance slit of the spectrometer is optically virtual. It is actually a diverging F/13 beam coming out from the polarization beam-splitter. This beam is collimated by a parabolic off-axis mirror (collimator~1). A single R2 echelle grating working in Littrow configuration (52.67 lines.mm$^{-1}$, 63.5° blaze angle) creates the principal dispersion. The dispersed beam is reflected again by the collimator 1 and forms an intermediate image. A small angle on the echelle grating allows the beam to pass on the side of the entrance beam. After reflecting on a second transfer collimator, the collimated beam reaches a dichroic plate, reflecting the UV range ([119:320] nm) and transmitting the visible part of the spectrum ([390:888] nm). 
The transmitted visible wavelengths are cross-dispersed using two LF5 prims. An unobscured 3-mirrors anastigmatic (TMA) telescope forms the spectrum on the detector, with a resolving power of 35000.
We propose a LiF grism as a cross-disperser for the UV range. A similar TMA telescope forms the spectrum on the UV detectors, with a resolving power of 25000.

	\subsection{Detectors}
Considering the whole spectral range covered by Arago, the spectrum will be spread onto three detectors: one for the visible part and two for the UV part.
\\

\begin{figure}[h]
\begin{center}
\fbox{\includegraphics[scale=0.25]{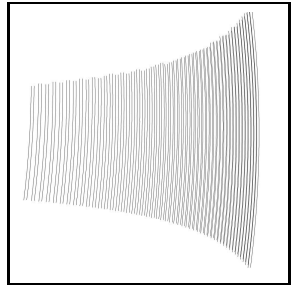}}
\caption{\label{detector_visible} Visible spectra onto the 30 $\times$ 30 mm$^2$ active area CCD detector.}
\end{center}
\end{figure}

For the visible part [390:888] nm, the choice of the detector is both determined by the observation strategy and the requirements applicable to the visible channel of the instrument. As we need an integration time of less than 20 min, a quantum efficiency above 0.7 over the spectrum, and a dark current of less than 0.01 e-.pix$^{-1}$.s$^{-1}$, we will use a thinned back-illuminated Double Depletion CCD with specific coating. It will be used in frame transfer mode to avoid shadowing effects and to optimize the observation time. The chosen CCD has a pixel pitch of 15 $\mu$m to meet the spectral resolution specification and an image size of $30 \times 30$ mm$^2$. An image of the spectra on the CCD is shown in Figure~\ref{detector_visible}.

\begin{figure}[h]
\begin{center}
\fbox{\includegraphics[scale=0.34]{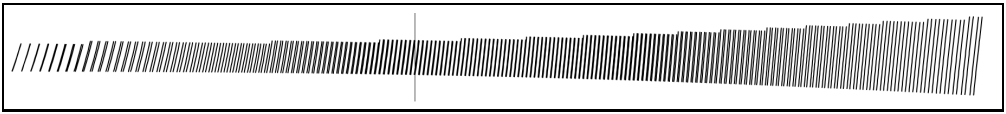}}
\caption{\label{uv} UV spectra onto the two MCP detectors. }
\end{center}
\end{figure}

The UV part of the spectrum has to be separated on two detectors to avoid the overcrossing of two diffraction orders. The first detector images the spectrum between 119 and 160~nm ($55 \times 14$~mm$^2$) and the other one between 160 and 320 nm ($55 \times 14$~mm$^2$). To achieve the requirement of 25000 in spectral resolution, the pixel needed pitch is 20~$\mu$m. Since the quantum efficiency of CCD or CMOS technologies in such wavelengths is too low, the choice was made to use image intensifier tubes with rectangular multichannel plates (MCP), each with a suitable photo-cathode material. The intensifier converts each incoming UV photons into an amplified electron pulse to be accelerated onto a phosphor screen, where it creates a spot of visible light. Fiber optical coupling transfers the visible photons to a serie of images sensors (CMOS/APS sensor). Figure~\ref{uv} shows an image of the spectra on the UV detectors.


\vspace{-2mm}
\section{Conclusions}
Thanks to a single instrument, providing high-resolution spectropolarimetry and precision full-Stokes polarization measurements, the Arago mission will allow breakthrough observations of stars and their environments. Particularly, the simultaneous spectropolarimetry observations in the UV [119:320] nm and visible [390:888] nm parts of the spectrum with the same instrument offer new horizons.


\begin{thebibliography}{}
\bibitem[Auriere et~al. (2003)]{auriere03}
{Auriere, M.} 2003, \textit{EAS Publications Series} 9, 105
\bibitem[Donati et al. (2006)]{donati06}
{Donati, J. F., Catala, C., Landstreet, J. D. and Petit, P.} 2006, \textit{Astronomical Society of the Pacific Confrence Series} 358, 362
\bibitem[Donati et al (1997)]{donati97}
{Donati, J. F., Semel, M., Carter, B. D., Rees, D. E. and Cameron, A. C.} 1997, 
\textit{Mon. Not. R. Astron. Soc. } 291
\bibitem[Pertenais et al (2014)]{pertenais14}
{Pertenais, M., Neiner, C., Pares, L., Petit, P., Snik, F., van Harten, G.} 
2014, \textit{Proc. SPIE } 9144
\bibitem[Snik et al (2012)]{snik12}
{Snik, F., van Harten, G., Navarro, R., Groot, P., Kaper, L. and de Wijn, A.} 
2012, \textit{Proc. SPIE } 8446
\bibitem[Tomczyk et al. (2010)]{tomczyk10}
{Tomczyk, S., Casini, R., de Wijn, A. G. and Nelson, P. G.} 
2010, \textit{Applied Optics} 49, 3580
\bibitem[del Toro Iniesta et al. (2000)]{deltoroiniesta00}
{del Toro Iniesta, J. C. and Collados, M.} 
2000, \textit{Applied Optics} 39, 1637


\end{thebibliography}
\end{document}